\documentstyle[preprint,eqsecnum,aps]{revtex}

\newcommand{\be}{\begin{equation}}
\newcommand{\ee}{\end{equation}}
\newcommand{\bea}{\begin{eqnarray}}
\newcommand{\eea}{\end{eqnarray}}

\begin{document}
\draft
\preprint{NCL93-TP17}
\title{Does scalar field collapse produce ``zero mass'' black holes?}
\author{Patrick  R.  Brady }
\address{
 Department of Physics, University of Newcastle, Newcastle upon Tyne NE1 7RU}
\date{December 14, 1993}
\maketitle
\begin{abstract}
An exact one parameter ($\alpha$) family of solutions representing scalar field
 collapse is presented.  These solutions exhibit a type of critical behaviour
 which has been discussed by Choptuik.  The three possible evolutions are
 outlined.  For supercritical evolution (when black holes form)  I show
that a quantity related to the mass of the black hole exhibits a power law
dependence on $\alpha$, for near critical evolution
$M \simeq |\alpha-\alpha_{\rm crit}|^{1/2}$.  Based on the properties of the
solution some comments are also made on conjectures of Choptuik.
\end{abstract}

\narrowtext

%\twocolumn

\narrowtext

The spherically symmetric collapse of a scalar field provides an ideal model in
which to investigate strong field dynamics of General Relativity.  This model
has therefore been considerably exploited both analytically~\cite{ref:1} and
numerically~\cite{ref:2,ref:3}.  In particular Christodoulou~\cite{ref:1} has
established many
general features of the solutions to the Einstein-scalar field equations, while
numerical studies~\cite{ref:2,ref:3} have provided some useful
insights into black hole formation.  Choptuik~\cite{ref:3} has produced some of
the
most fascinating results thus far.  His analysis of the problem highlights
several intriguing non-linear phenomena and motivates the present work.  This
Letter describes some analytic results of an investigation of the
collapse problem, and a search for a theoretical explanation of the behaviour
discovered by Choptuik~\cite{ref:3}.

In order to make some progress I assume that the collapse is self-similar: i.e.
there exists a vector field, $\xi$, such that the spacetime metric $g$,
satisfies ${\cal
L}_{\xi} g = 2 g$ where $\cal L_{\xi}$ denotes the Lie derivative with respect
to $\xi$.  Homothetic gravitational collapse has been studied in great
detail~\cite{ref:4},  and many examples of naked singularities fall into
this class.  In this case it allows me to obtain a one parameter family of
solutions representing scalar field collapse.
This family exhibits some of the properties discussed by
Choptuik~\cite{ref:3} and Christodoulou~\cite{ref:1}.  Specifically the
solutions exhibit  critical
behaviour:  if $\alpha$ is the parameter which characterises the solutions then
when $\alpha > 0$ the scalar field collapses, interacts and disperses leaving
behind nothing but flat space.  The exactly critical case is $\alpha_{\rm crit}
= 0$ and corresponds to a spacetime which is asymptotically flat but which
contains a null, scalar-curvature singularity at $r=0$.  The remaining case,
$\alpha < 0$, corresponds to black hole formation.  Unfortunately, as occurs
in other spacetimes with a homothetic symmetry, the black hole eventually grows
to infinite mass.  It would therefore seem difficult to obtain any information
about
the scaling law for black hole mass discovered by Choptuik~\cite{ref:3}.
However, when the quantity $M$ (defined in Eq.~(\ref{eqn:exp}), roughly, the
ratio of
the mass at the apparent horizon, to the mass function at past null infinity)
is calculated one obtains the remarkable relation
	\be
	M = 2 |\alpha|^{1/2} (1-4\alpha)^{1/2}\: .
	\ee
Since $\alpha_{\rm crit} \equiv 0$ in this solution,  evolutions which are
sufficiently close to critical exhibit an approximate power law in $\alpha -
\alpha_{\rm crit}$.  The exponent is significantly different from that found
by Choptuik however.

I do {\em not} claim that the solution below is the universal strong field
solution which Choptuik~\cite{ref:3} conjectures may exist.  The goal of this
work is only to present the family of solutions in which some of the strong
field features can be explicitly seen, a more general analysis of homothetic
scalar field collapse will appear elsewhere.

The solution is most easily understood in terms of null coordinates on the
radial two space.  Therefore I write
	\be
	ds^2 = -2e^{2\sigma} du dv + r^2
	d\Omega^2 \; , \label{eqn:2}
	\ee
where $\sigma$ and $r$ are functions of both
$u$ and $v$, and $d\Omega^2$ is the line element on the unit sphere.  The
Einstein-scalar field equations are then
	\bea
	(r^2)_{,uv} &=& - e^{2\sigma} \; , \label{eqn:3}\\
	2 \sigma_{,uv} - \frac{2 r_{,u} r_{,v}}{r^2} &=&
	\frac{e^{2\sigma}}{r^2} - 2 \phi_{,u} \phi_{,v} \; , \label{eqn:4}\\
	r_{,vv} - 2 \sigma_{,v} r_{,v} &=& - r (\phi_{,v})^2 \; ,
	\label{eqn:5}\\
	r_{,uu} - 2 \sigma_{,u} r_{,u} &=& - r (\phi_{,u})^2 \; ,
	\label{eqn:6}
	\eea
and the wave equation for the scalar field $\phi$
	\be
	2\phi_{,uv} + (\ln r^2)_{,v} \phi_{,u}
	+ (\ln r^2)_{,u} \phi_{,v} =0 \; .  \label{eqn:7}
	\ee
Here a
comma ($,$) denotes a partial derivative.  Notice that these equations are
invariant in form under reparametrizations of the null coordinates $u$ and $v$.
Due to the spherical symmetry of the problem it is possible to introduce a
local mass function $m(u,v)$ by
	\be
	1 - \frac{2m(u,v)}{r} := 2 g^{uv} r_{,u} r_{,v} \; .
	\label{eqn:mass}
	\ee
This mass function agrees with both the ADM and Bondi masses in the appropriate
limits.

A solution to Eqs.~(\ref{eqn:3})-(\ref{eqn:7}) is
	\bea
	r^2 &=& \alpha v^2 + \beta u^2 - uv \; , \label{eqn:8}\\
	e^{2\sigma} &=& 1 \; , \label{eqn:9}
	\eea
where the scalar field is given by
	\be
	\phi = \left\{
	\begin{array}{lr}
	\pm \frac{1}{2}
	\ln \left| \frac{ \displaystyle
	2\alpha v - u(1+\sqrt{1-4\alpha\beta})}
	{ \displaystyle
	2\alpha v - u(1-\sqrt{1-4\alpha\beta})}\right|
	&\ \ \alpha \ne 0 \\
	\pm \frac{1}{2} \ln \left|\beta -\frac{\displaystyle v}{
	\displaystyle u}\right|
	&\ \ \alpha = 0
        \end{array}\right. \label{eqn:11}
        	\ee
It is assumed that
$\phi \equiv 0$ for $v<0$,  where $v$ is standard advanced time.  The mass
function as defined by (\ref{eqn:mass}) is
	\be
	m(u,v) =  -uv(1-4\alpha\beta)/4r \label{eqn:mass1}\; ,
	\ee
and the homothetic vector field is
	\be
	\xi = u\frac{\partial}{\partial u} + v\frac{\partial}{\partial v}\; .
	\ee
It was the presence of this symmetry which enabled the integration of the
equations initially.

In order to discuss the problem of collapse we can set $\beta = 1$ without loss
of generality.  \footnote{By adjusting $\beta$ one obtains the time reverse of
the solutions discussed below, that is one obtains white hole spacetimes,
along with some other self-similar cosmological solutions.}
The solutions now fall into three distinct classes depending on
the value of the parameter $\alpha$.  In the language used by
Choptuik~\cite{ref:3}
these are subcritical ($0<\alpha < 1/4$), critical ($\alpha = 0$) and
supercritical ($\alpha <0$).  We consider the influx of scalar field to be
turned on at the advanced time $v=0$ so that to the past of this surface the
spacetime is Minkowskian and the spacetime metric is therefore $C^1$ at this
surface.  Moreover notice when $\alpha = 1/4$ the solution is simply flat
space.

Let us now consider each of the three classes in turn:
\newline
(i) {\it Subcritical} ($0 <\alpha <1/4$):  Fig.  1 shows
a conformal diagram for a typical spacetime in this class.  The scalar field is
initially dispersed, collapsing from past null infinity and interacting
gravitationally.  However, the collapse never forms a black hole.  The mass
function is identically zero on $u=0$ (as it is along $v=0$) and there is no
flux of scalar field across this surface.  In fact the only non-vanishing
component of stress-energy
at $u=0$ is
	\be
	T_{uu} = \frac{v^2 (1-4\alpha\beta)}{4r^4}.
	\ee
Thus the spacetime is flat for $u>0$.

The solution represents a scalar field collapse in which the
gravitational interaction is never strong enough to form a black hole.  The
existence of such solutions has been established previously by
Christodoulou~\cite{ref:1}.
\newline
(ii) {\it Critical} ($\alpha =0$):  In this spacetime
the initial configuration is again collapse of scalar field from past null
infinity with a flat, vacuum interior region.  The collapse now proceeds to a
singularity at $r=0$.  Using Eq.~(\ref{eqn:8}) with $\alpha =0$ we see that
this corresponds to $u=0$,  and the singularity is therefore null (see Fig.~2).
One easily checks that the mass function (\ref{eqn:mass1}) tends to zero as
$u\rightarrow 0$, however the curvature diverges there: $|\Psi_2| = m/r^3 \sim
u^{-3/2}$ along an ingoing null ray.

It is also straightforward to check that all outgoing null rays reach infinity;
there is no apparent horizon in this spacetime. In particular
	\be
	\lim_{u\rightarrow 0 } \left[ 1 - \frac{2m}{r}\right] = \frac{1}{2}\; .
	\label{eqn:apph}
	\ee
More will be said about this in a moment.  Finally I wish to emphasise that
the spacetime does not violate the
strong cosmic censorship hypothesis~\cite{ref:5} since no observer can see
the singularity without actually reaching it.  Of course by waiting long
enough an observer can always ``see'' regions of arbitrarily large
curvature.
\newline
(iii) {\it Supercritical} ($\alpha <0$):
Fig.~3  shows the global structure of these spacetimes.
In the past they behave the same as in the previous two cases, however there is
now an apparent horizon surrounding a spacelike $r=0$ singularity.
Unfortunately, the mass of the resulting black hole is infinite and the entire
spacetime ultimately gets trapped. These spacetimes are unphysical
(as black hole spacetimes) unless the influx of scalar field from past infinity
is turned off at some finite advanced time. Something more can however be said
if one examines the mass function along the apparent horizon of the spacetime.

The locus of the apparent horizon is $ r=2m $, or in terms of the coordinates
	\be
	u = 2\alpha v \: .
	\ee
Substituting this into (\ref{eqn:mass1}) we find that
	\be
	m_{\rm AH} = \frac{v}{2}\sqrt{-\alpha}\sqrt{1-4\alpha} \; ,
	\ee
where the subscript $AH$ indicates that this is along the apparent horizon.
Let us define a dimensionless quantity related to the mass of this black hole
as follows
	\be
	M = \frac{m_{AH}}{m|_{u=-\infty}} = \frac{2 |\alpha|^{1/2}}
	{\sqrt{1-4\alpha}}  \label{eqn:exp} \; .
	\ee
We notice that $M$ exhibits a power-law dependence on the
parameter $\alpha$.  This rings of the observation which Choptuik~\cite{ref:3}
and others~\cite{ref:6} have made about the masses of the black holes formed in
nearly critical collapse.  It also seems reasonable that the quantity defined
in (\ref{eqn:exp}) should in fact agree with that measured by Choptuik as
$v\rightarrow \infty$.  There is a discrepancy though, the exponent has been
calculated numerically to be approximately $.37$ and not $1/2$ as found here.
Whether this has any significance is the subject of further investigations.

As a final comment on the black hole mass in near critical evolutions
let me mention a possible explanation for the difference in exponents above.
For a nearly critical evolution one may imagine turning off the influx from
infinity at some finite advanced time $v_1 > 0$, say, then  the mass of the
resulting black hole will be
	\be
	 m|_{v\rightarrow \infty} = \frac{v_1}{2}
	  \sqrt{-\alpha}\sqrt{1-4\alpha\beta}\:
	 +\Delta m \; ,
	 \ee
where  $2\Delta m$ is the amount by which the apparent horizon increases in
radius during the evolution from $v_1$ to $\infty$.  For black holes formed in
nearly critical collapse it is entirely possible that this ``correction'' term
dominates,  since much of the scalar field remains outside the apparent horizon
up to the advanced time $v_1$.  It would therefore seem worthwhile to estimate
$\Delta m$.  To date I have been unable to do so.

I wish also to point out the connection between this solution and two
conjectures which have been made about the near critical evolution,  and black
hole formation.  In previous numerical searches~\cite{ref:2} there appeared to
be a lower bound on the masses of black holes formed in supercritical
evolution,  however the more recent surveys of scalar field
collapse~\cite{ref:3,ref:6} have been capable of far greater resolution and the
results led to the following two conjectures:
(1)~Black holes initially form with infinitesmal mass.  (2)~The exactly
critical evolution is a ``zero mass'' black hole.

The above solution [Eqs.~(\ref{eqn:8})-(\ref{eqn:11})] clearly lends support to
the first of these conjectures. Provided one cuts off the influx from infinity
at some finite advanced time, it then seems that one can make the mass of the
resulting black hole as small as one likes.

On the second conjecture, we have seen that the evolution does not really form
a black hole at all.  The spacetime contains no apparent horizon,  not even at
$r=0$ as shown by (\ref{eqn:apph}),  while the mass function actually becomes
infinite on future null infinity.  Thus, it certainly could not be interpreted
as a zero-mass black hole.  More generally I suggest that the exactly critical
evolution will stand distinct from either the supercritical or subcritical
cases.

It is a pleasure to thank Viqar Husain, Erik Martinez and Dario
Nunez for stimulating discussions during a short visit to the
University of Alberta.  I also acknowledge financial support
from the Scientific and Engineering Research Council of Great Britain. After
the completion of this work Werner Israel drew my attention
to M.D. Roberts,  Gen.
Relativ. Grav.  {\bf 21}, 907 (1989) where the solution
(\ref{eqn:8})-(\ref{eqn:11}) was previously discussed in the context of
counter-examples to cosmic censorship.

\begin{figure}

\caption{  This is the diagram for a subcritical evolution ($0<\alpha,1/4$).
The spacetime is topologically trivial.  Indicated are the lines $v=0$ where
the scalar field influx is turned on, and $u=0$ where the spacetime beyond
which the spacetime returns to flat space.  The shaded region is where the
scalar field is non-zero.  ${\cal M}_4$ indicates regions which are Minkowski.}
\end{figure}

\begin{figure}
 \caption{  The exactly critical solution ($\alpha = 0$).  The initial
configuration is similar to the subcritical case;  $v=0$ is where the scalar
field influx is turned on with flat space (${\cal M}_4$) to the past of it.
The shaded region is where the scalar field is non-zero.  Indicated is the null
scalar curvature singularity.  Notice that all $r={\rm const.}$ lines terminate
at $P$,  which is future timelike infinity.}
\end{figure}

\begin{figure}
 \caption{  Supercritical evolutions ($\alpha <0$) lead to the entire spacetime
becoming trapped.  The initial configuration is once again very similar to the
subcritical case; $v=0$ is where the scalar field influx is turned on.  AH is
the apparent horizon,  notice that it is spacelike everywhere and ``surrounds''
the spacelike $r=0$ singularity.}
\end{figure}


\begin{references}

\bibitem{ref:1} D. Christodoulou, Commun. Math. Phys. {\bf 105}, 337 (1986);
                {\bf 106}, 587 (1986); {\bf 587} (1986); {\bf 109}, 591 (1987);
                {\bf 109}, 613 (1987).

\bibitem{ref:2} M.W. Choptuik, in {\em Frontiers in Numerical Relativity}, C.R.
                Evans, L.S. Finn and D.W. Hobill (eds.), Cambridge University
                Press,  Cambridge (1989);  A. Goldwirth and T. Piran, Phys.
		Rev.
                D{\bf 36}, 3575 (1987); R. G\'omez and J. Winicour,  in {\em
                 Frontiers in Numerical Relativity}, C.R.
                Evans, L.S. Finn and D.W. Hobill (eds.), Cambridge University
                Press,  Cambridge (1989);   R. G\'omez, R.A. Isaacson
                and J. Winicour,  J. Comp. Phys. {\bf 98}, 11 (1992).

\bibitem{ref:3} M.W. Choptuik, Phys. Rev. Letters {\bf 70}, 9 (1993).



\bibitem{ref:4} M.E. Cahill and A.H. Taub,  Commun. Math. Phys. {\bf 21},1
                (1971);  D.M. Eardley and L. Smarr,  Phys. Rev. D{\bf 19}, 2239
                (1979);  D. Christodoulou,  Commun. Math. Phys. {\bf 93},
                171 (1984); R.P.A.C. Newman, Class. Quantum Grav. {\bf 3}, 527
                (1986);
                A. Ori and T. Piran, Phys. Rev. D{\bf 42}, 1068 (1990);
                R.N. Henriksen and K. Patel, Gen. Relativ. Grav. {\bf 23}, 527
                (1991);  K.  Lake and T. Zannias,  Phys. Rev. D{\bf 41}, 3866
                (1990).

\bibitem{ref:5} R. Penrose,  in {\em An Einstein Centenary Survey}, S.W.
                 Hawking and W. Israel (eds.),  Cambridge University Press,
                  Cambridge (1979).

\bibitem{ref:6} C. Gundlach, R. Price and J. Pullin, {\em Late-time behaviour
of
stellar collapse and explosions: II.  Nonlinear evolution},  preprint
NSF-ITP-93-85,  gr-qc/9307010.

\end{references}
\end{document}